\documentclass[prd,twocolumn,groupedaddress,showpacs,nofootinbib]{revtex4-1}
\usepackage{graphicx}
\usepackage{dcolumn}
\usepackage{amssymb}
\usepackage{mathrsfs}
\usepackage{amsmath}
\usepackage{epsfig}
\usepackage[dvips]{color}
\usepackage{hhline}

\begin{document}

\title{High-scale baryogenesis with testable neutron-antineutron oscillation and dark matter}

\author{Pei-Hong Gu$^{1}_{}$}
\email{peihong.gu@sjtu.edu.cn}

\author{Utpal Sarkar${}^{2}$}
\email{utpal.sarkar.prl@gmail.com}

\affiliation{$^{1}_{}$Department of Physics and Astronomy, Shanghai Jiao Tong University, 800 Dongchuan Road, Shanghai 200240, China\\
${}^{2}$Physics Department, Indian Institute of Technology, Kharagpur 721302, India}

\begin{abstract}

We propose a new scenario for predicting a one-loop neutron-antineutron oscillation at a testable level, meanwhile, realizing a thermal or inflationary baryogenesis at a very high scale. Besides the standard model content, this scenario involves two real singlet scalars with very heavy masses, two color-triplet and iso-singlet scalars at the TeV scale, as well as a Majorana singlet fermion for a dark matter candidate.

\end{abstract}

\pacs{98.80.Cq, 95.35.+d, 14.20.Dh}

\maketitle

\section{Introduction}

It is well known that in the $SU(3)_c^{}\times SU(2)_L^{}\times U(1)_Y^{}$ standard model (SM), an $SU(2)_L^{}$
global anomaly violates the baryon $(B)$ and lepton $(L)$ numbers by an equal amount \cite{thooft1976}. Due to a sphaleron solution \cite{krs1985}, this anomalous process can become fast during the period $ 100\,\textrm{GeV} \lesssim T \lesssim 10^{12}_{}\,\textrm{GeV}$. Consequently, a $B-L$ asymmetry rather than a $B+L$ asymmetry can survive from such $B-L$ conserving but $B+L$ violating sphaleron processes. This means a successful baryogenesis mechanism above the electroweak scale should firstly generate a $B-L$ asymmetry which is composed of a pure $B$ asymmetry, a pure $L$
asymmetry or any unequal $B$ and $L$ asymmetries. The sphaleron processes then can convert this $B-L$ asymmetry to a $B$ asymmetry and a $L$ asymmetry. For example, the $B-L$ asymmetry is a $L$ asymmetry in the leptogenesis scenario where a $L$ asymmetry can be produced by certain lepton-number-violating interactions \cite{fy1986}. Alternatively, we can consider some baryon-number-violating interactions to generate a $B$ asymmetry for the required $B-L$ asymmetry. These baryon number violation may result in a proton decay \cite{pss1983,gs2011,bm2012,takhistov2014,gms2016} or a neutron-antineutron ($n-\bar{n}$) oscillation \cite{gms2016,ns2001,wn2010,bmn2006,bmn2006-2,gu2007,gs2011-2,bbv2011,bm2012-2,bc2012,dm2015,phillips2014}. 

On the other hand, the massive and mixing neutrinos have been confirmed by various neutrino oscillation experiments \cite{patrignani2016}. The precise cosmology further constrains the neutrino masses in a sub-eV range \cite{patrignani2016}. The most popular scheme for generating the tiny neutrino masses is to consider the seesaw mechanism at tree level \cite{minkowski1977,yanagida1979,grs1979,ms1980,sv1980}. In such seesaw extension of the SM, the cosmic $B$ asymmetry can be understood via the leptogenesis mechanism. An interesting variation of the canonical seesaw model is to introduce an inert Higgs doublet and then realize the seesaw at one-loop order \cite{ma2006,ma2015}. This radiative seesaw scenario, where the leptogenesis is still available while the inert Higgs doublet provides a dark matter (DM) candidate, inspire us to consider the possibility of a one-loop induced $n-\bar{n}$ oscillation.

In this paper we shall propose a new scenario with a high-scale baryogenesis and a one-loop $n-\bar{n}$ oscillation. Specifically, we shall extend the SM by two real singlet scalars, two color-triplet and iso-singlet scalars with same electric charges but different $B$ numbers, as well as a Majorana singlet fermion without any $B$ numbers. Our model will respect a softly broken $B$ number and an exactly conserved $Z_2^{}$ discrete symmetry. The baryon-number-violating decays of the real scalars into the colored scalars, followed by the baryon-number-conserving decays of the colored scalars into the SM quarks and the Majorana fermion, can realize a thermal or inflationary baryogenesis. The interactions responsible for this baryogenesis can also contribute a $n-\bar{n}$ oscillation at one-loop order. This $n-\bar{n}$ oscillation can be observed in the future experiments if the colored scalars and the Majorana fermion are close to the TeV scale. Meanwhile, the Majorana fermion as a DM particle can be verified in the DM direct detection experiments.

\section{The model}

Besides the quarks and Higgs scalar in the SM,
\begin{eqnarray}
&&d_{R}^{}(3,2,-\!\begin{array}{c}\frac{1}{3}\end{array}\!)(+\!\begin{array}{c}\frac{1}{3}\end{array}\!)\,,
~~u_{R}^{}(3,2,+\!\begin{array}{c}\frac{2}{3}\end{array}\!)(+\!\begin{array}{c}\frac{1}{3}\end{array}\!)\,,\nonumber\\
[1mm]
&&q_{L}^{}(3,2,+\!\begin{array}{c}\frac{1}{6}\end{array}\!)(+\!\begin{array}{c}\frac{1}{3}\end{array}\!)=\left[\begin{array}{c} u^{}_{L} \\
[1mm] d_{L}^{}\end{array}\right]\,,\nonumber\\
[1mm]
&&\phi(1,2,+\!\begin{array}{c}\frac{1}{2}\end{array}\!)(0)=\left[\begin{array}{c} \phi^{+}_{} \\
[1mm] \phi_{}^{0}\end{array}\right]\,,
\end{eqnarray}
our scenario will involve the following five fields,
\begin{eqnarray}
&&\chi_R^{}(1,1,0)(0)\,,~~\delta(3,1,-\!\begin{array}{c}\frac{1}{3}\end{array}\!)(-\!\begin{array}{c}\frac{2}{3}\end{array}\!)\,,
~~\xi(3,1,-\!\begin{array}{c}\frac{1}{3}\end{array}\!)(-\!\begin{array}{c}\frac{1}{3}\end{array}\!)\,,\nonumber\\
[1mm]
&&\sigma_a^{}(1,1,0)(0)~~(a=1,2)\,,
\end{eqnarray}
with $\chi_R^{}$ being a fermion while $\delta$, $\xi$ and $\sigma_{1,2}^{}$ being four scalars. Here the first brackets following the fields describe the transformations under the $SU(3)_c^{}\times SU(2)_L^{}\times U(1)^{}_{Y}$ gauge groups, while the second brackets are the $B$ numbers. We also impose a $Z_2^{}$ discrete symmetry under which the fields transform as
\begin{eqnarray}
\label{z2}
\!\!\!\!\!\!\!\!(\textrm{SM}\,,\,\delta)\stackrel{Z_2^{}}{\longrightarrow}(\textrm{SM}\,,\,\delta)\,,~(\chi_R^{}\,,\,\xi\,,\,\sigma) \stackrel{Z_2^{}}{\longrightarrow}-(\chi_R^{}\,,\,\xi\,,\,\sigma)\,.
\end{eqnarray}

In our model, the $B$ number is only permitted to be softly broken, meanwhile, the $Z_2^{}$ symmetry is required to be exactly conserved. Under these conditions, the following terms relevant to our demonstration can be allowed,
\begin{eqnarray}
\label{lag}
\mathcal{L}&\supset&-\frac{1}{2}M_{\sigma}^2\sigma^2_{}-(\mu_\delta^2+ \lambda^{}_{\delta\phi}\phi^\dagger_{}\phi)\delta^\dagger_{}\delta-(\mu_\xi^2+ \lambda^{}_{\xi\phi}\phi^\dagger_{}\phi)\xi^\dagger_{}\xi\nonumber\\
[1mm]
&&-\left[\frac{1}{2}m_{\chi}^{}\bar{\chi}_{R}^{}\chi_{R}^c+\rho\sigma\delta^\dagger_{}\xi+ y\xi^{\dagger}_{} \bar{d}_{R}^{c}\chi_{R}^{}+f\delta\bar{d}_{R}^{c} u_{R}^{}  \right.\nonumber\\
[1mm]
&&\left.+\frac{1}{2}h\delta\bar{q}_{L}^c i\tau_2^{}q_{L}^{}+\textrm{H.c.}\right]\,,
\end{eqnarray}
where the indices are not shown. Note the two parameters $\rho_{1,2}^{}$ always can have a relative phase,
\begin{eqnarray}
\label{phase}
\alpha=\arg\left(\frac{\rho_2^{}}{\rho_1^{}}\right)\neq 0\,,
\end{eqnarray}
to violate the CP.

The $Z_2^{}$ discrete symmetry will not be broken at any scales. This means the real singlet scalars $\sigma$ will not acquire any nonzero vacuum expectation values (VEVs). As for the colored scalars $\delta$ and $\xi$, their masses should be
\begin{eqnarray}
m_{\delta}^2=\mu_\delta^2+ \lambda^{}_{\delta\phi}\langle\phi\rangle^2_{}>0\,,~~
m_{\xi}^2=\mu_\xi^2+ \lambda^{}_{\xi\phi}\langle\phi\rangle^2_{}>0\,.
\end{eqnarray}
Without loss of generality, we choose the Majorana mass of the singlet fermion $\chi_R^{}$ to be real so that we can define a Majorana fermion, i.e.
\begin{eqnarray}
\chi=\chi_R^{}+\chi_R^c=\chi^c_{}~~\textrm{for}~~m_\chi^{}=m_\chi^\ast\,.
\end{eqnarray}

We would like to emphasize that under the $Z_2^{}$ symmetry (\ref{z2}), the soft $B$ number violation can only originate from the trilinear couplings among the three scalars $\sigma$, $\delta$ and $\xi$, i.e. the $\rho$-term in Eq. (\ref{lag}). As we will demonstrate in the following, such baryon-number-violating terms with the CP-violation (\ref{phase}) play an essential role for realizing a thermal or inflationary baryogenesis.

\section{Dark matter}

In our model, the Majorana fermion $\chi$ can remain stable to serve as a DM particle. The DM relic density is determined by the annihilation of the DM fermion $\chi$ into the down-type quark pairs through the t-channel exchange of the colored scalar $\xi$. Meanwhile, the s-channel exchange of the colored scalar $\xi$ can mediate a DM scattering off nuclei. For simplicity, we assume that the colored scalar $\xi$ to be
much heavier than the DM fermion so that the DM annihilation and scattering can be well described by the effective operator as below,
\begin{eqnarray}
\mathcal{L}&\supset&\frac{y_i^{}y_j^\ast}{m_\xi^2}\bar{d}_{Ri}^{}\chi_R^{c} \bar{\chi}_R^{c}d_{Rj}^{}=\frac{y_i^{}y_j^\ast}{8m_\xi^2}\bar{d}_{i}^{}\gamma^\mu_{}(1-\gamma_5^{})d_{j}^{}\bar{\chi}\gamma_\mu^{}\gamma_5^{}\chi\,.\nonumber\\
&&
\end{eqnarray}

We now calculate the annihilation cross section up to the $p$-wave contribution \cite{kt1990},
\begin{eqnarray}
\langle\sigma v_{\textrm{rel}}^{}\rangle&=&\frac{\sum_{i,j=d,s,b}^{}|y_i^{}|^2_{}|y_j^{}|^2_{}}{4\,\pi}\frac{m_\chi^2}{m_\xi^4}v_{\textrm{rel}}^{2}\,.
\end{eqnarray}
By taking the freeze-out temperature $T\simeq m_\chi^{}/20$ and then the relative velocity $v_{\textrm{rel}}^{2}\simeq 6 T/m_\chi^{}\simeq 0.3$, we can estimate the annihilation cross section to be
\begin{eqnarray}
\langle\sigma v_{\textrm{rel}}^{}\rangle
&=&0.1\,\textrm{pb}\left(\frac{\sum_{i,j=d,s,b}^{}|y_i^{}|^2_{}|y_j^{}|^2_{}}{1}\right)\nonumber\\
[1mm]
&&\times \left(\frac{m_\chi^{}}{100\,\textrm{GeV}}\right)^2_{}
\left(\frac{1\,\textrm{TeV}}{m_\xi^{}}\right)^4_{}\,.
\end{eqnarray}
For a proper choice of the masses and couplings, the above cross section can arrive at a desired value to account for the DM relic density \cite{patrignani2016}. In the following we will simply assume the colored scalar $\xi$ does not sizeably couple to the second and third generations of quarks. As an example, we take
\begin{eqnarray}
&&m_\chi^{}=100\,\textrm{GeV}\,,~~m_\xi^{}=1\,\textrm{TeV}\,,~~y_d^{}=1\gg y_{s,b}^{}\nonumber\\
[1mm]
&&\Rightarrow \langle\sigma v_{\textrm{rel}}^{}\rangle =0.1\,\textrm{pb}\,.
\end{eqnarray}

As the colored scalar $\xi$ has no sizeable couplings to the strange and bottom quarks, the spin-dependent DM-nucleus scattering cross section at zero momentum transfer can be given by \cite{jkg1996}
\begin{eqnarray}
\sigma_0^{}=\frac{4|y_d^{}|^4_{}}{\pi}\frac{\mu_A^2}{m_\xi^4}[\Delta d_{}^{(p)}\langle S_p^{}\rangle+\Delta d_{}^{(n)}\langle S_n^{}\rangle]^2_{}
\frac{J+1}{J}\,.
\end{eqnarray}
Here $\mu_A^{}=m_A^{}m_\xi^{}/(m_A^{}+m_\xi^{})$ is the DM-nucleus reduced mass with $m_A^{}$ being the target nuclear mass, $\Delta d_{}^{(p)}=-0.38$ \cite{jkg1996} and $\Delta d_{}^{(n)}=0.77$ \cite{jkg1996} respectively are the down-quark matrix element in a proton and a neutron, $\langle S_p^{}\rangle$ is the expectation value of the spin content of the proton group in the nucleus, and similarly for $\langle S_n^{}\rangle$, while $J$ is the total nucleus spin. As an example, we consider the $^{129}\textrm{Xe}$ nucleus with $\langle S_p^{}\rangle=0.010$, $\langle S_n^{}\rangle=0.329$, $J=1/2$ \cite{kmgs2012}, as well as the $^{131}\textrm{Xe}$ nucleus with $\langle S_p^{}\rangle=-0.009$, $\langle S_n^{}\rangle=-0.272$, $J=3/2$ \cite{kmgs2012}. We find $\sigma_0^{}=1.5\times 10^{-36}_{}\,\textrm{cm}^2_{}$ for $^{129}\textrm{Xe}$ and $\sigma_0^{}=6.0\times 10^{-37}_{}\,\textrm{cm}^2_{}$ for $^{131}\textrm{Xe}$. By making a normalization procedure $\sigma_N^{}=\sigma_0^{}\mu_N^2/\mu_A^{2}$, we can obtain a DM-nucleon scattering cross section $\sigma_N^{}=8\times 10^{-41}_{}\,\textrm{cm}^2_{}$ for $^{129}\textrm{Xe}$ and $\sigma_N^{}=3\times 10^{-41}_{}\,\textrm{cm}^2_{}$ for $^{131}\textrm{Xe}$, which are close to the current limits and sensitive to the future experiments \cite{fu2016}. In this numerical analysis, we have defined the DM-nucleon reduced mass $\mu_N^{}=m_N^{}m_\xi^{}/(m_N^{}+m_\xi^{})$ with $m_N^{}$ being the nucleon mass.

\section{Neutron-antineutron oscillation}

\begin{figure}
\vspace{5.5cm} \epsfig{file=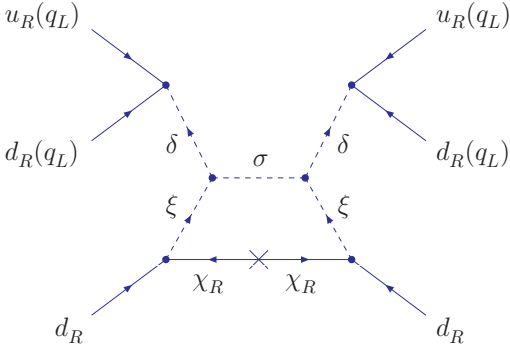, bbllx=5cm, bblly=6.0cm,
bburx=15cm, bbury=16cm, width=7cm, height=7cm, angle=0,
clip=0} \vspace{-7.75cm} \caption{\label{nnbar} The one-loop diagram for neutron-antineutron oscillation.}
\end{figure}

At low energy, we can integrate out the real scalars $\sigma$, the colored scalars $\delta$ and $\xi$, as well as the Majorana fermion $\chi$. Consequently, as shown in Fig. \ref{nnbar}, we can obtain the dimension-9 operators as below,
\begin{eqnarray}
\label{dim9}
\mathcal{L}&\supset&-\frac{\kappa}{\Lambda^5_{R}}\bar{d}_{R}^{c}d_R^{}\bar{u}_R^{c}d_R^{}\bar{u}_R^{c}d_R^{}
-\frac{\kappa}{\Lambda^5_{L}}\bar{d}_R^{c}d_R^{}\bar{u}_L^{c}d_L^{}\bar{u}_L^{c}d_L^{}\nonumber\\
[1mm]
&&+\textrm{H.c.}~~\textrm{with}\nonumber\\
[1mm]
\frac{\kappa_{ijklmn}^{}}{\Lambda^5_{R}}&=&\frac{y_i^{}y_j^{}f_{kl}^{}f_{mn}^{}}{4\pi^2_{}}\frac{\rho_a^2m_\chi^{}\left[1-\frac{m_\chi^2}{m_\xi^2-m_\chi^2}
\ln\left(\frac{m_\xi^2}{m_\chi^2}\right)\right]}{M_{\sigma_a^{}}^2m_\delta^4(m_\xi^2-m_\chi^2)}\nonumber\\
[1mm]
&\simeq&\frac{y_i^{}y_j^{}f_{kl}^{}f_{mn}^{}}{4\pi^2_{}}\frac{\rho_a^2m_\chi^{}}{M_{\sigma_a^{}}^2 m_\xi^2 m_\delta^4 }~~\textrm{for}~~m_\xi^2\gg m_\chi^2\,,\nonumber\\
[1mm]
\frac{\kappa_{ijklmn}^{}}{\Lambda^5_{L}}&=&\frac{y_i^{}y_j^{}h_{kl}^{}h_{mn}^{}}{4\pi^2_{}}\frac{\rho_a^2m_\chi^{}\left[1-\frac{m_\chi^2}{m_\xi^2-m_\chi^2}
\ln\left(\frac{m_\xi^2}{m_\chi^2}\right)\right]}{M_{\sigma_a^{}}^2m_\delta^4(m_\xi^2-m_\chi^2)}\nonumber\\
[1mm]
&\simeq&\frac{y_i^{}y_j^{}h_{kl}^{}h_{mn}^{}}{4\pi^2_{}}\frac{\rho_a^2m_\chi^{}}{M_{\sigma_a^{}}^2 m_\xi^2 m_\delta^4 }~~\textrm{for}~~m_\xi^2\gg m_\chi^2\,.
\end{eqnarray}

The above operators, which violate the $B$ number by two units, can result in a $n-\bar{n}$ transition. The related $n-\bar{n}$ mixing strength is calculated by \cite{phillips2014}
\begin{eqnarray}
\!\!\!\!\!\!\!\!\!\!\delta m_{n-\bar{n}}^{}\!&\sim&\! \Lambda_{\textrm{QCD}}^6 G_{n-\bar{n} }^{}\frac{}{}\nonumber\\
[1mm]
\!\!\!\!\!\!\!\!\!\!\!&=&\!3.4\times 10^{-5}_{}\,\textrm{GeV}\left(\frac{\Lambda_{\textrm{QCD}}^{}}{180\,\textrm{MeV}}\right)^6_{}\left(\frac{G_{n-\bar{n}}^{}}{\textrm{GeV}^{-5}_{}}\right),
\end{eqnarray}
where the parameter $G_{n-\bar{n}}^{}$ is
\begin{eqnarray}
G_{n-\bar{n}}^{}&=& G_{R}^{}+G_{L}^{}~~\textrm{with}\nonumber\\
[1mm]
G_{R}^{}&=&\frac{\kappa_{ddudud}^{}}{\Lambda_R^5}=\frac{y_d^2 f_{ud}^2}{4\pi^2_{}}\frac{m_\chi^{}}{ m_\xi^2 m_\delta^4 }\sum_{a}^{}\frac{\rho_a^2}{M_{\sigma_a^{}}^2}\nonumber\\
[1mm]
&=&10^{-28}_{}\,\textrm{GeV}^{-5}_{}\times \left(\frac{y_d^{}}{1}\right)^2_{}\left(\frac{f_{ud}^{}}{4.5\times 10^{-3}_{}}\right)^2_{}\nonumber\\
[1mm]
&&
\times \left(\frac{m_\chi^{}}{100\,\textrm{GeV}}\right)\left(\frac{1\,\textrm{TeV}}{m_\xi^{}}\right)^2_{}\left(\frac{1\,\textrm{TeV}}{m_\delta^{}}\right)^4_{}
\nonumber\\
[1mm]
&&\times
\sum_a^{}\left(\frac{\rho_a^{}/M_{\sigma_a^{}}^{}}{10^{-3}_{}}\right)^2_{}\,,\nonumber\\
[1mm]
G_{L}^{}&=&\frac{\kappa_{ddudud}^{}}{\Lambda_L^5}=\frac{y_d^2 h_{ud}^2}{4\pi^2_{}}\frac{m_\chi^{}}{m_\xi^2 m_\delta^4}\sum_{a}^{}\frac{\rho_a^2}{M_{\sigma_a^{}}^2}\nonumber\\
[1mm]
&=&10^{-28}_{}\,\textrm{GeV}^{-5}_{}\times \left(\frac{y_d^{}}{1}\right)^2_{}\left(\frac{h_{ud}^{}}{4.5\times 10^{-3}_{}}\right)^2_{}\nonumber\\
[1mm]
&&\times \left(\frac{m_\chi^{}}{100\,\textrm{GeV}}\right)\left(\frac{1\,\textrm{TeV}}{m_\xi^{}}\right)^2_{}
\left(\frac{1\,\textrm{TeV}}{m_\delta^{}}\right)^4_{}\nonumber\\
[1mm]
&&\times \sum_a^{}\left(\frac{\rho_a^{}/M_{\sigma_a^{}}^{}}{10^{-3}_{}}\right)^2_{}\,.
\end{eqnarray}
The $n-\bar{n}$ oscillation with the lifetime,
\begin{eqnarray}
\tau_{n-\bar{n}}^{}=\frac{1}{\delta m_{n-\bar{n}}^{}}\approx 2\times 10^8_{}\,\textrm{sec} \left(\frac{10^{-28}_{}\,\textrm{GeV}^{-5}_{}}{G_{n-\bar{n}}^{}}\right)\,,
\end{eqnarray}
can arrive at a testable level \cite{phillips2014}.

\section{Baryogenesis}

\begin{figure*}
\vspace{5.5cm} \epsfig{file=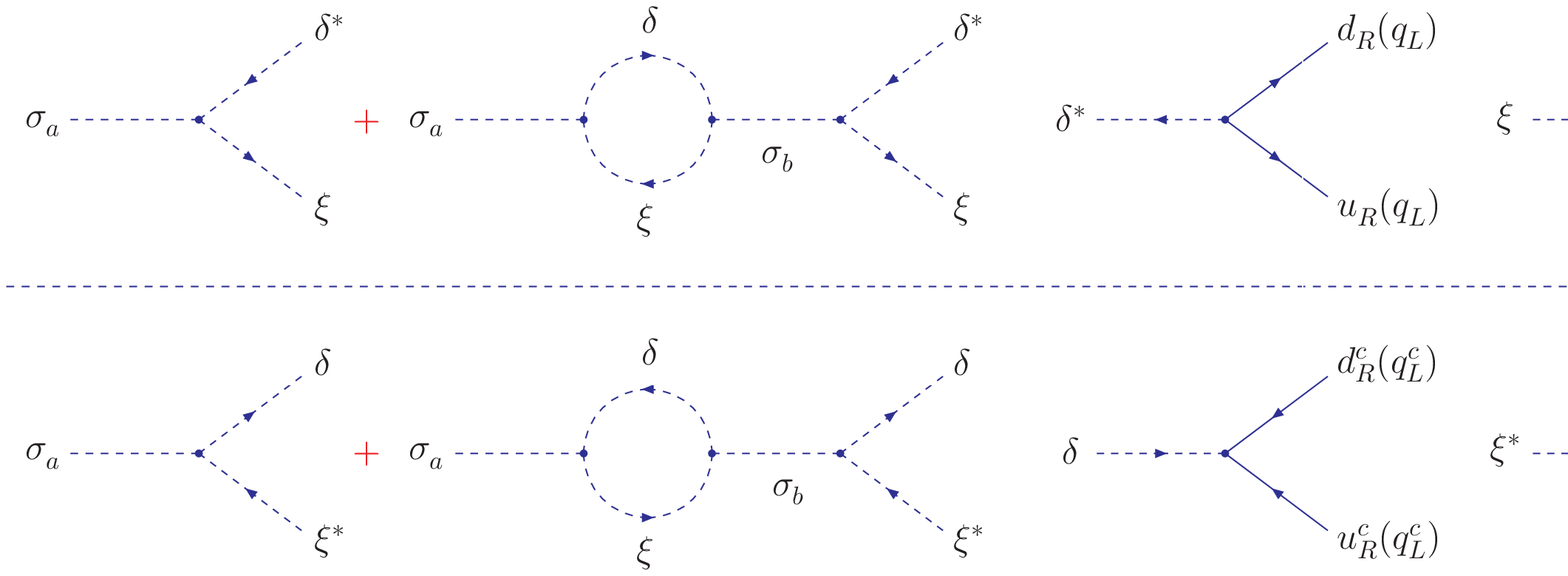, bbllx=7.5cm, bblly=6.0cm,
bburx=17.5cm, bbury=16cm, width=7cm, height=7cm, angle=0,
clip=0} \vspace{-6.75cm} \caption{\label{decay} The real scalars ($\sigma_a^{}$) decay into the colored scalars ($\delta$ and $\xi$) which subsequently decay into the SM quarks ($d_R^{}$, $u_R^{}$ and $q_L^{}$) as well as the singlet fermion ($\chi_R^{}$). }
\end{figure*}

As shown in Fig. \ref{decay}, the baryon-number-violating decays of the real scalars $\sigma_a^{}$ can generate a $B$ asymmetry stored in the colored scalar pairs $(\delta\,,\delta^\ast_{})$ and $(\xi\,,\xi^\ast_{})$. Subsequently, through the baryon-number-conserving decays of the colored scalars, the SM quarks can acquire a $B$ asymmetry, which is equivalent to a $B-L$ asymmetry in the absence of a $L$ asymmetry \footnote{In the Majorana neutrino case, we assume the $B$ asymmetry will be produced after the lepton-number-violating interactions for generating the neutrino mass matrix $m_\nu^{}$ decouple at a very high temperature \cite{fy1990},
\begin{eqnarray}
T=10^{12}_{}\,\textrm{GeV}\left[\frac{0.04\,\textrm{eV}^2_{}}{\textrm{Tr}(m_\nu^\dagger m_\nu^{})}\right]\,.
\end{eqnarray}
For the Dirac neutrinos, their mass generation conserves the $L$ number so that it will not affect the produced $B$ asymmetry at all.}. This $B-L$ asymmetry can survive from the sphaleron processes and then contribute a $B$ asymmetry in the present universe.

We calculate the width in the real scalar decays at tree level,
\begin{eqnarray}
\Gamma_{\sigma_a^{}}^{}&=&\Gamma(\sigma_a^{}\rightarrow \delta^\ast_{}+\xi)+\Gamma(\sigma_a^{}\rightarrow \delta+\xi^\ast_{})=\frac{3}{8\pi}\frac{|\rho_{a}^{}|^2_{}}{M_{\sigma_a^{}}^{}}\,,\nonumber\\
&&
\end{eqnarray}
and the CP asymmetry at one-loop order,
\begin{eqnarray}
\varepsilon_{\sigma_a^{}}^{}&=&\frac{\Gamma(\sigma_a^{}\rightarrow \delta^\ast_{}+\xi)-\Gamma(\sigma_a^{}\rightarrow \delta+\xi^\ast_{})}{\Gamma_{\sigma_a^{}}^{}}\nonumber\\
[2mm]
&=&\frac{3}{8\pi}\frac{\textrm{Im}(\rho_{a}^{2}\rho_{b}^{\ast 2})}
{|\rho_{a}^{}|^2_{}}\frac{1}{M_{\sigma_b^{}}^{2}-M_{\sigma_a^{}}^{2}}\,.
\end{eqnarray}

As an example, we assume the real scalar $\sigma_{1}^{}$ to be much lighter than the other one $\sigma_{2}^{}$. The final $B$ asymmetry then should mainly come from the $\sigma_{1}^{}$ decays. For a numerical estimation, we define \cite{kt1990}
\begin{eqnarray}
\label{rwidth}
K&=&\frac{\Gamma_{\sigma_{1}^{}}^{}}{2H(T)}\left|_{T=M_{\sigma_{1}^{}}^{}}^{}\right.\,,
\end{eqnarray}
where $H(T)$ is the Hubble constant,
\begin{eqnarray}
H=\left(\frac{8\pi^{3}_{}g_{\ast}^{}}{90}\right)^{\frac{1}{2}}_{}
\frac{T^{2}_{}}{M_{\textrm{Pl}}^{}}\,,
\end{eqnarray}
with $g_{\ast}^{}$ being the relativistic degrees of freedom during the baryogenesis epoch. In the weak washout region, the final $B$ asymmetry can be simply described by \cite{kt1990}
\begin{eqnarray}
\eta_B^{}=\frac{n_B^{}}{s}\simeq \frac{28}{79}\times \frac{\varepsilon_{\sigma_{1}^{}}^{}}{g_\ast^{}}~~\textrm{for}~~K\ll 1\,.
\end{eqnarray}
Here $n_B^{}$ and $s$, respectively, are the $B$ number density and the entropy density, while the factor $\frac{28}{79}$ is a sphaleron transfer coefficient. By fixing $g_\ast^{}=120.5$ (the SM fields plus two color-triplet and iso-singlet scalars as well as one singlet fermions) and inputting,
\begin{eqnarray}
\label{pchoice}
M_{\sigma_{1}^{}}^{}&=&10^3_{}\,|\rho_{1}^{}|=10^{12}_{}\,\textrm{GeV}\,,\nonumber\\
[1mm]
M_{\sigma_{2}^{}}^{}&=&10^3_{}\,|\rho_{2}^{}|=10^{13}_{}\,\textrm{GeV}\,,
\end{eqnarray}
we read
\begin{eqnarray}
K=0.04\,,~~\varepsilon_{\sigma_{1}^{}}^{}=3.3\times 10^{-8}_{}\left(\frac{\sin 2\alpha}{0.28}\right)\,.
\end{eqnarray}
The $B$ asymmetry then can arrive at an expected value \cite{patrignani2016},
\begin{eqnarray}
\label{basy}
\eta_B^{}= 10^{-10}_{}\left(\frac{\sin 2\alpha}{0.28}\right)\,.
\end{eqnarray}

Alternatively, the scalar $\sigma_{1}^{}$ may play the role of an inflaton \cite{klw2014}. In this case, the final $B$ asymmetry should be \cite{kt1990}
\begin{eqnarray}
\eta_B^{}=\frac{28}{79}\varepsilon_{\sigma_{1}^{}}^{} T_{\textrm{RH}}^{}/M_{\sigma_{1}}^{}\,,
\end{eqnarray}
with $T_{\textrm{RH}}^{}$ being the reheating temperature \cite{kt1990},
\begin{eqnarray}
T_{\textrm{RH}}^{}&\equiv &T (t=\Gamma_{\sigma_{1}^{}}^{-1})\nonumber\\
[1mm]
&=&\left(\frac{90}{8\pi^{3}_{}g_{\ast}^{}}\right)^{\frac{1}{4}}_{}
\sqrt{\frac{3M_{\textrm{Pl}}^{} |\rho_1^{}|^2_{}}{16\pi M_{\sigma_{1}^{}}^{}}}\,.
\end{eqnarray}
Replacing the parameter choice (\ref{pchoice}) by
\begin{eqnarray}
M_{\sigma_{1}^{}}^{}&=&10^3_{}\,|\rho_{1}^{}|=1.5\times 10^{13}_{}\,\textrm{GeV}\,,\nonumber\\
[1mm]
M_{\sigma_{2}^{}}^{}&=&10^3_{}\,|\rho_{2}^{}|=1.5\times 10^{14}_{}\,\textrm{GeV}\,,
\end{eqnarray}
we find
\begin{eqnarray}
T_{\textrm{RH}}^{}&=&7.7\times  10^{11}_{}\,\textrm{GeV}\,,\nonumber\\
[1mm]
\varepsilon_{\sigma_{1}^{}}^{}&=&5.6\times 10^{-9}_{}\left(\frac{\sin 2\alpha}{0.047}\right)\,,\nonumber\\
[1mm]
\eta_B^{}&=& 10^{-10}_{}\left(\frac{\sin 2\alpha}{0.047}\right)\,.
\end{eqnarray}

One may worry about the produced $B$ asymmetry will be erased at low energies because the dimension-9 operators violate the $B$ number by two units and hence lead to some baryon-number-violating processes. Usually ones estimate the rate of these processes by
\begin{eqnarray}
\label{estimation}
G_{n-\bar{n}}^{2}T^{11}_{}<H(T)&\Longrightarrow& T\lesssim 10^{4}_{}\,\textrm{GeV}\nonumber\\
&&\textrm{for}~~G_{n-\bar{n}}^{}=10^{-28}_{}\,\textrm{GeV}^{-5}_{}\,,~~
\end{eqnarray}
and then conclude no $B$ asymmetry can survive above the temperature $T\sim 10^{4}_{}\,\textrm{GeV}$ if the $n-\bar{n}$ oscillation arrives at a testable level. In our model, the effective dimension-9 operators are induced by integrating out the scalars $\sigma$, $\delta$ and $\xi$. However, the scalars $\delta$ and $\xi$ are at the TeV scale, i.e. their masses are lighter than the crucial temperature $T\sim 10^{4}_{}\,\textrm{GeV}$. So, the estimation (\ref{estimation}) is not consistent with the present scenario. Actually, in our model, the cubic terms among the three scalars $\sigma$, $\delta$ and $\xi$ provide the unique source of the $B$ number violation. After this $B$ number violation is decoupled, no other baryon-number-violating processes can keep in equilibrium to wash out the produced $B$ asymmetry.

\section{Summary}

In this paper we have demonstrated an interesting scenario that a high-scale baryogenesis can be consistent with a testable $n-\bar{n}$ oscillation. Our model contains two real singlet scalars, two colored scalars and a Majorana singlet fermion, in addition to the SM content. We impose a softly broken $B$ number and an exactly conserved $Z_2^{}$ discrete symmetry. The baryon-number-violating decays of the real scalars and then the baryon-number-conserving decays of the colored scalars can generate a $B$ asymmetry stored in the SM quarks. The interactions responsible for this baryogenesis can also result in a $n-\bar{n}$ oscillation at one-loop order. Such radiative $n-\bar{n}$ oscillation can arrive at a testable level even if the real scalars are much heavier than the colored scalars and the Majorana fermion which are all close to the TeV scale. Meanwhile, the Majorana fermion as a DM particle is sensitive to the DM direct detection. The real scalars even can drive an inflation.

\textbf{Acknowledgement}: P.H.G. was supported by the National Natural Science Foundation of China under Grant No. 11675100, the Recruitment Program for Young Professionals under Grant No. 15Z127060004, the Shanghai Jiao Tong University under Grant No. WF220407201, the Shanghai Laboratory for Particle Physics and Cosmology, and the Key Laboratory for Particle Physics, Astrophysics and Cosmology, Ministry of Education.

\end{document}